\documentclass[aps,prbtwocolumn,superscriptaddress,citeautoscript,amssymb,reprint,showpacs,floatfix,longbibliography]{revtex4-2}
\usepackage[ansinew]{inputenc}
\usepackage[T1]{fontenc}
\usepackage{float}
\usepackage{amsmath}
\usepackage[normalem]{ulem}
\usepackage{mathptmx}
\usepackage{mathrsfs}
\usepackage[english]{babel}
\usepackage{graphicx}
\usepackage[usenames,dvipsnames]{color}
\usepackage{natbib}
\usepackage{etoolbox}
\usepackage{makecell}
\usepackage[thinspace,thinqspace]{SIunits}
\usepackage{nicefrac}
\usepackage[colorlinks=true,linkcolor=blue,citecolor=blue,urlcolor=blue]{hyperref}
\newcommand{\CRA}{CeRh$_2$As$_2$}
\newcommand{\Tc}{$T_{\textrm{c}}$}
\newcommand{\To}{$T_{\textrm{0}}$}

\newcommand{\Tceven}{$T_{\textrm{c}}^{\textrm{SC1}}$}
\newcommand{\Tcodd}{$T_{\textrm{c}}^{\textrm{SC2}}$}

\makeatletter
\patchcmd{\@bibitem}{\@biblabel}{\@biblabel}{}{}
\patchcmd{\@bibitem}{``}{\ignorespaces}{}{}
\patchcmd{\@bibitem}{''}{\ignorespaces}{}{}
\makeatother

\begin{document}
\author{K.~Semeniuk}
\thanks{These authors contributed equally.}
\affiliation{Max Planck Institute for Chemical Physics of Solids, 01187 Dresden, Germany}

\author{M.~Pfeiffer}
\thanks{These authors contributed equally.}
\affiliation{Institute for Solid State and Materials Physics, TU Dresden University of Technology, 01062 Dresden, Germany}
\affiliation{Max Planck Institute for Chemical Physics of Solids, 01187 Dresden, Germany}

\author{J.~F.~Landaeta}
\affiliation{Institute for Solid State and Materials Physics, TU Dresden University of Technology, 01062 Dresden, Germany}
\affiliation{Max Planck Institute for Chemical Physics of Solids, 01187 Dresden, Germany}

\author{M.~Nicklas}
\affiliation{Max Planck Institute for Chemical Physics of Solids, 01187 Dresden, Germany}

\author{C.~Geibel}
\affiliation{Max Planck Institute for Chemical Physics of Solids, 01187 Dresden, Germany}

\author{M.~Brando}
\affiliation{Max Planck Institute for Chemical Physics of Solids, 01187 Dresden, Germany}

\author{S.~Khim}
\affiliation{Max Planck Institute for Chemical Physics of Solids, 01187 Dresden, Germany}

\author{E.~Hassinger}
\thanks{correspondence should be addressed to konstantin.semeniuk@cpfs.mpg.de and elena.hassinger@tu-dresden.de}
\affiliation{Institute for Solid State and Materials Physics, TU Dresden University of Technology, 01062 Dresden, Germany}
\affiliation{Max Planck Institute for Chemical Physics of Solids, 01187 Dresden, Germany}

\title{Exposing the odd-parity superconductivity in CeRh$_2$As$_2$ with hydrostatic pressure}
\date{\today}

\begin{abstract}
Odd-parity superconductivity is a fundamentally interesting but rare state of matter with a potential for applications in topological quantum computing. Crystals with staggered locally noncentrosymmetric structures have been proposed as platforms where a magnetic field can induce a transition between even- and odd-parity superconducting (SC) states. The strongly correlated superconductor \CRA, with the critical temperature $T_{\mathrm{c}}\approx0.4$\,K, is likely the first example material showing such a phase transition, which occurs at the magnetic field $\mu_{0}H^{*}=4$\,T applied along the crystallographic $c$ axis. \CRA\ also undergoes a phase transition of an unknown origin at $T_{0}=0.5$\,K. By subjecting \CRA\ to hydrostatic pressure and mapping the resultant changes to the SC phase diagrams we investigated how the lattice compression and changes to the electronic correlations affect the stability and relative balance of the two SC states. The abnormally high in-plane upper critical field becomes even higher close to a quantum critical point of the \To\ order. Remarkably, the SC phase-switching field $H^{*}$ is drastically reduced under pressure, dropping to 0.3\,T at 2.7\,GPa. This result signals an apparent strengthening of the local noncentrosymmetricity and forecasts a possible stabilization of the putative odd-parity state down to zero field, hitherto not considered by theoretical models.
\end{abstract}
\maketitle

\textit{Introduction.}---Broken inversion symmetry in crystals causes momentum-dependent spin polarization, as exemplified by the Rashba effect~\cite{bihlmayer2015}. In noncentrosymmetric superconductors, coupling of the momentum and spin degrees of freedom prohibits labeling the Cooper-pair wave function as a purely spin-singlet or spin-triplet one --- the state is a mix of both~\cite{gorkov2001}. However, two noncentrosymmetric lattices can be combined into a globally centrosymmetric superstructure, with the constituents related via the inversion symmetry. The superconducting (SC) state must then have a definite parity and a magnetic-field-induced switching from an even parity state to an odd-parity state is expected when the polarization of the spins due to the sizable spin-orbit coupling remains preserved locally and stabilizes the odd-parity state~\cite{yoshida2012}.

The two-phase superconductivity in \CRA\ may be the only known practical realization of this phenomenon~\cite{khim2021}. The material crystallizes in a non-symmorphic CaBe$_2$Ge$_2$-type structure (space group 129, P4/nmm)~\cite{madar1987}, and in accordance with the above-mentioned principle, its unit cell has two non-centrosymmetric Ce sites of opposite polarity, related to each other via inversion symmetry.

\CRA\ is an unconventional superconductor with the critical temperature $T_{\mathrm{c}}\approx$~0.3--0.4\,K, varying with sample quality~\cite{khim2021,semeniuk2023}. Applying a magnetic field $H_{||}$ parallel to the crystallographic $c$ axis reveals two peculiar aspects: 1) At ${\mu_{0}H^{*}=4}$\,T, a transition from a low-field SC state (SC1) to a high-field state (SC2) takes place; 2) The upper critical field $H_{\mathrm{c2}}^{||}$ reaches 15\,T~\cite{khim2021,landaeta2022}, which is unusually high for the given \Tc. The SC2 state rapidly vanishes as the field is rotated, and for a field $H_{\perp}$ perpendicular to the $c$ axis, only the SC1 phase exists with the upper critical field $H_{\mathrm{c2}}^{\perp}$ of 2\,T~\cite{landaeta2022}. Based on a reasonable agreement of these observations with the earlier predictions~\cite{yoshida2012}, SC1 and SC2 have been interpreted, respectively, as the states of the even and odd parity of the order parameter~\cite{khim2021}. The conjecture is further supported by recent theoretical works~\cite{moeckli2021b,cavanagh2022,fischer2023}.

At a temperature $T_{0}\approx0.5$\,K, \CRA\ enters a state "Phase~I" of currently unknown microscopic origin~\cite{hafner2022,mishra2022,semeniuk2023}. At a pressure $P_{0}\approx0.5$\,GPa, Phase~I vanishes in a quantum critical point (QCP), which gives rise to the quantum critical fluctuations that apparently drive the SC pairing and lead to the extreme quasiparticle masses and non-Fermi-liquid behavior in the system~\cite{khim2021,pfeiffer2024}. Additionally, recent nuclear magnetic and quadrupolar resonance studies detected local magnetic fields below 0.25\,K and proposed an antiferromagnetic (AFM) order~\cite{kibune2022,kitagawa2022,ogata2023}, but it is currently unclear whether it is concomitant with \To\ or \Tc\ transitions, or exists independently from the other states~\cite{chajewski2024}.

The alternating Rashba-type spin-orbit coupling due to the local inversion symmetry breaking and strong electronic correlations of a heavy-fermion system are thought to be the key ingredients for realizing the parity transition. The former favors comparable magnitudes of the SC pairing potentials of the two phases. The latter allows the purely orbitally-limited SC2 state to survive at higher fields than the paramagnetically-limited SC1 state. The low orbital limiting field of isostructural weakly-correlated LaRh$_{2}$As$_{2}$ can then explain why this compound hosts only one SC phase~\cite{landaeta2022b}. 

We used hydrostatic pressure to investigate how tuning both the strength of correlations and the local non-centrosymmetricity affects the unusual superconductivity in \CRA. While an independent study~\cite{siddiquee2023} reported a discontinuity in $T_{\mathrm{c}}(P)$ at 2.5\,GPa, hinting at a pressure-induced transition to a different SC state, this result was not reproduced in our previous work~\cite{pfeiffer2024}, and there are strong indications that the anomaly was caused by a sudden loss of pressure hydrostaticity, rendering the results for $P\geq2.5$\,GPa questionable. The study also suffered from a sizable uncertainty in critical temperatures and fields at high pressures, making it impossible to draw solid conclusions regarding the effect of pressure on the relative balance between SC1 and SC2 (we analyze the results of Ref.~\cite{siddiquee2023} in more detail in Supplemental Material [SM]).

In this letter, we demonstrate that the in-plane upper critical field is enhanced and the shape of the $H_{\mathrm{c2}}^{\perp}(T)$ curve becomes anomalous close to $P_{0}$, in agreement with the presence of the QCP~\cite{pfeiffer2024}. More importantly, our study reveals a pronounced and interesting effect of pressure on the multi-phase superconductivity: while both SC1 and SC2 occupy a smaller area in the $H_{\parallel}-T$ space at high pressures, SC2 becomes more favored with respect to SC1 (contrary to Ref.~\cite{siddiquee2023}). This is accompanied by a dramatic reduction of the phase-switching field $H^*$, signifying a major enhancement of the staggered Rashba interaction.

\begin{figure}
    \includegraphics[width=\columnwidth]{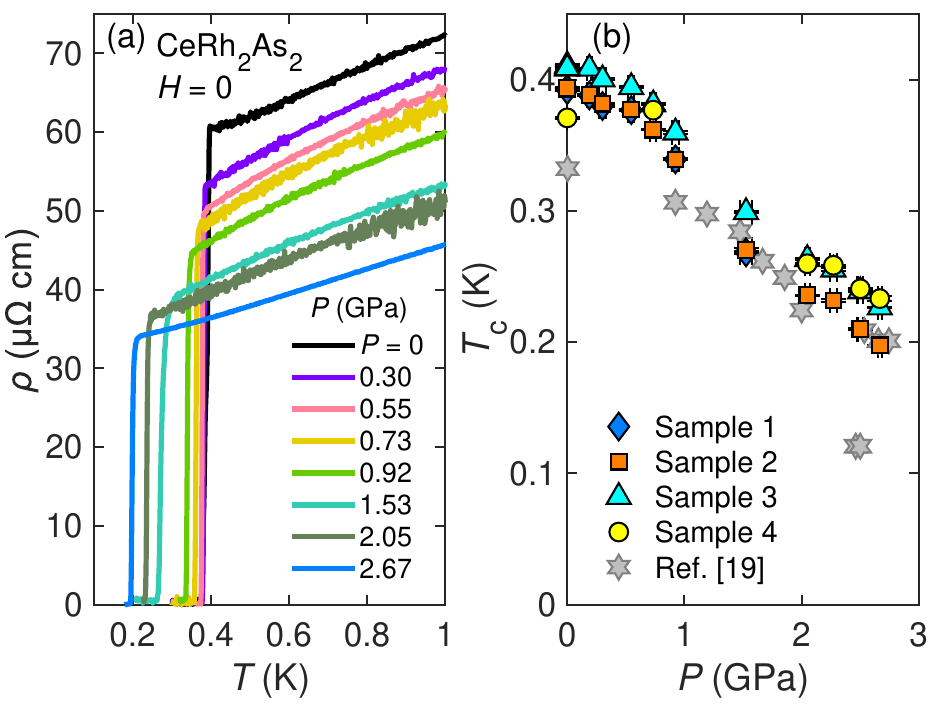}
    \caption{Effect of pressure on the superconducting transition of \CRA. (a) Resistivity ($\rho$) of \CRA\ (Sample 2) against temperature ($T$) at different pressures. (b) Pressure ($P$) dependence of the transition temperature \Tc\ for different samples, including data from an independent study~\cite{siddiquee2023}. \Tc\ is defined as the temperature at which $d\rho/dT$ peaks. Differences between the samples are explained in Supplemental Material.
    }
    \label{fig:ZeroFieldData}
\end{figure}

\textit{Results.}---In Fig.~\ref{fig:ZeroFieldData}(a) we show how pressure affects the SC transition seen in $\rho(T)$ at zero magnetic field, and in Fig.~\ref{fig:ZeroFieldData}(b) we plot resistive $T_{\mathrm{c}}(P)$ for multiple samples studied. Besides a $\sim30$\,mK spread of \Tc, all of the measured samples show consistent pressure dependence, which agrees with the one found in our previous study~\cite{pfeiffer2024}. At ambient pressure, the system is close to the peak of the SC dome in the $T$--$P$ space, and \Tc\ effectively stays at $\sim0.4$\,K up to about $0.7$\,GPa. Subsequently, \Tc\ steadily decreases down to $\sim0.2$\,K at 2.67\,GPa, the highest achieved pressure. Compared, to the independent high-pressure study of \CRA~\cite{siddiquee2023}, the overall rate of suppression of \Tc\ is similar, yet the sharp dip near 2.5\,GPa is definitely absent for all our samples, indicating that the dip was almost certainly caused by the loss of hydrostaticity~\cite{pfeiffer2024}.

\begin{figure}
     \includegraphics[width=\columnwidth]{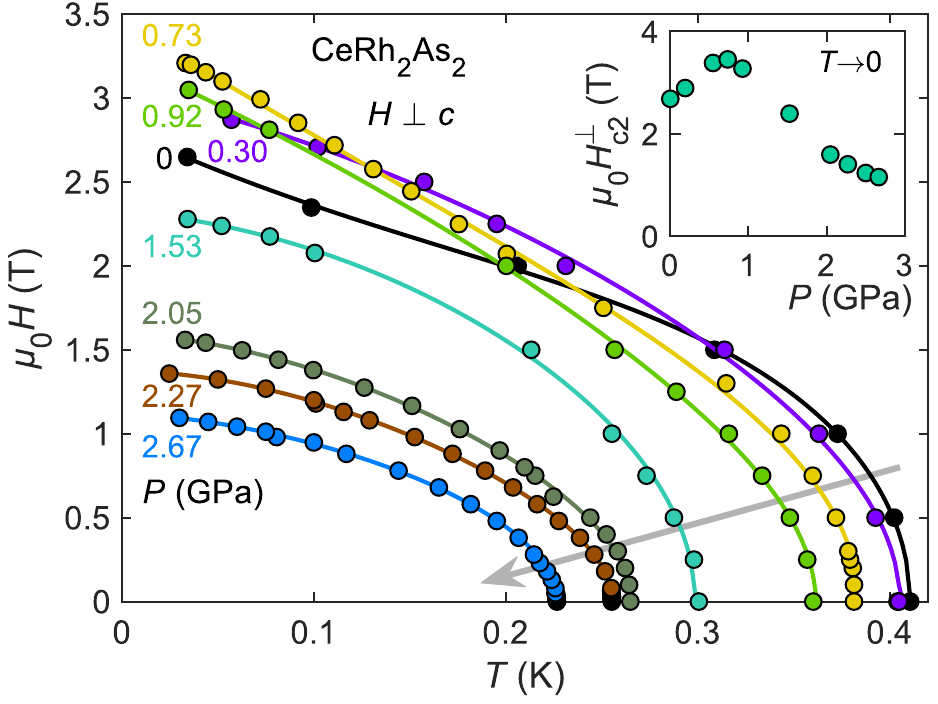}
     \caption{Superconductivity in \CRA\ under pressure ($P$) in the in-plane magnetic field $H\perp c$. (a) Field-temperature ($\mu_{0}H$--$T$) phase diagrams (Sample 3). The solid lines are guides for the eye. The grey arrow indicates the change for increasing pressure. Inset: the zero-temperature upper critical field against pressure.}
     \label{fig:PhaseDiagrams_ab}
\end{figure}

The in-plane upper critical field curves, $H_{\mathrm{c2}}^{\perp}(T)$, are shown in Fig.~\ref{fig:PhaseDiagrams_ab} for different pressures. A selection of $\rho(T)$ and $\rho(H)$ data behind the displayed phase diagrams is given in SM. For $T\ll T_{\mathrm{c}}$, $H_{\mathrm{c2}}^{\perp}$ initially increases with pressure and peaks between 0.55 and 0.73\,GPa. At these pressures, the $H_{\mathrm{c2}}^{\perp}(T)$ curve follows a rather atypical linear increase in the low-temperature region without signs of saturation down to 30\,mK. At higher pressures, $H_{\mathrm{c2}}^{\perp}$ decreases in the entire temperature range, and the saturating form expected from a strongly Pauli-limited superconductor is recovered by 1.5\,GPa.

\begin{figure}
     \includegraphics[width=\columnwidth]{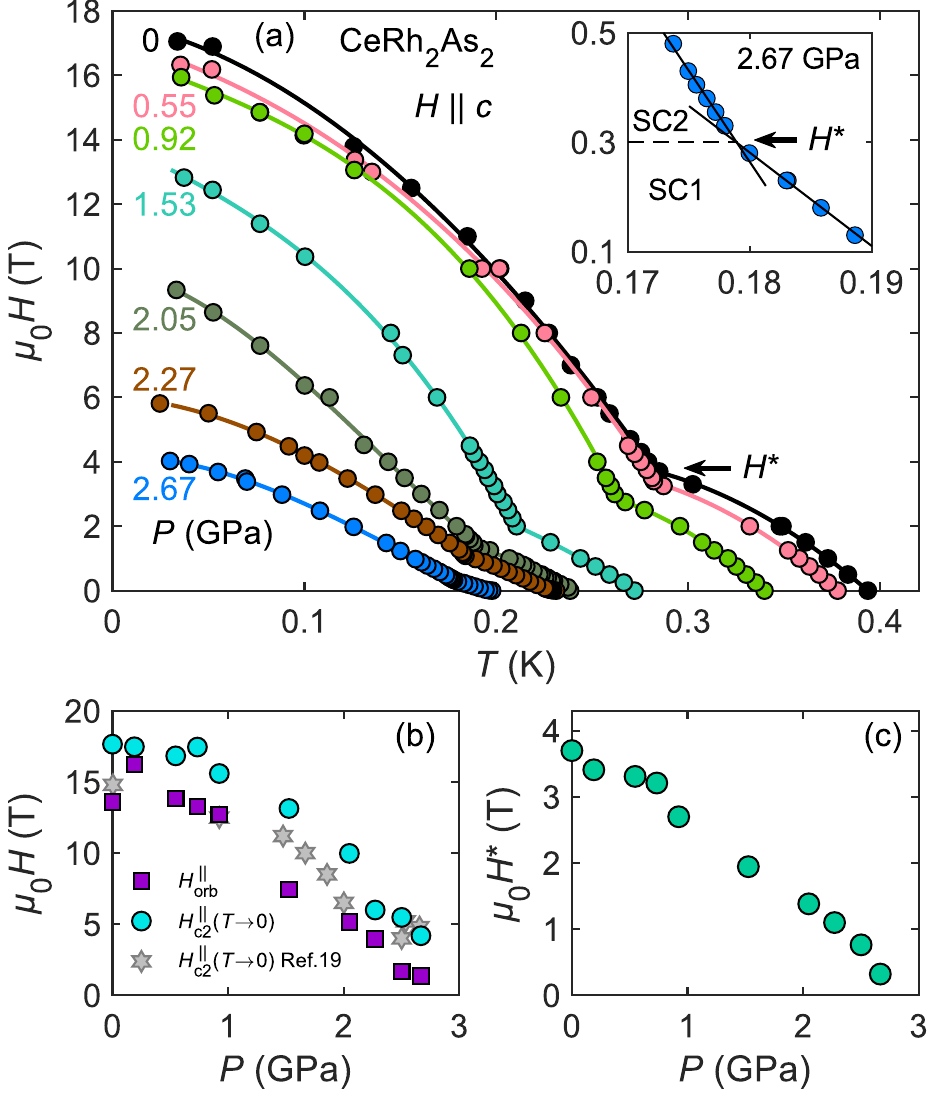}
     \caption{Superconductivity in \CRA\ under pressure ($P$) in the out-of-plane magnetic field $H\parallel c$. (a) Field-temperature ($\mu_{0}H$--$T$) phase diagrams (Samples 1 and 2). The solid lines are guides for the eye. The transition between the two superconducting phases SC1 and SC2 occurs at a magnetic field $H^{*}$ and is seen as a kink in the critical field curve (marked for $P=0$). Inset: the SC1-SC2 transition at 2.67\,GPa. The straight trend lines emphasize the slope change. The dashed line marks a hypothetical SC1-SC2 phase boundary. (b) Pressure dependence of the orbital-limiting upper critical field ($H_{\mathrm{orb}}^{\parallel}$), estimated from the slopes of $H_{\mathrm{c2}}^{\parallel}(T)$ curves at \Tc\, as well as the zero-temperature limit of the upper critical field $H_{\mathrm{c2}}^{||}$. Data from an independent study~\cite{siddiquee2023} are included for comparison. (c) Pressure dependence of $H^{*}$.}
     \label{fig:PhaseDiagrams_c}
\end{figure}

The upper critical field curves for the out-of-plane magnetic field, $H_{\mathrm{c2}}^{||}(T)$, are depicted in Fig.~\ref{fig:PhaseDiagrams_c}(a). The superconductivity is monotonically suppressed by pressure at all temperatures. In the zero-temperature limit, $H_{\mathrm{c2}}^{||}$ is reduced from 18\,T to 4\,T across the full pressure range [Fig.~\ref{fig:PhaseDiagrams_c}(b)]. The reduction of low-temperature $H_{\mathrm{c2}}^{||}$ in the SC2 state mimics the decrease of the orbital limiting field $H_{\mathrm{orb}}^{||}$, estimated from the slope of the critical field curve near \Tc, in the SC1 state, but there exists a ${\sim4}$\,T offset between the two quantities. The SC1-SC2 transition appears as a pronounced kink in $H_{\mathrm{c2}}^{||}(T)$. The kink shifts down in both field and temperature with pressure, and can be tracked all the way to 2.67\,GPa, where $\mu_{0}H^{*}=0.3$\,T, indicating that the two SC phases are still present [inset of Fig.~\ref{fig:PhaseDiagrams_c}(a)]. Figure~\ref{fig:PhaseDiagrams_c}(c) shows a steady decrease of $H^{*}$ with pressure.

\textit{Discussion.}---Whereas the anomalously high Pauli limit in \CRA\ has been mainly attributed to the enhancement caused by the Rashba interaction~\cite{khim2021}, electronic correlations may also play a crucial role~\cite{nogaki2022}. In particular, strong fluctuations in the vicinity of the Phase~I QCP could explain the non-trivial pressure dependence of the SC1 upper critical field for $H\perp c$ and the large Pauli limit. The Chandrasekhar-Clogston limit is violated in a number of other strongly-correlated superconductors, which do not exhibit a pronounced inversion symmetry breaking at the relevant sites~\cite{squire2023,miclea2006,kittaka2016,thomas1996}. In these cases, the large Pauli limit has been attributed to a strong coupling of electrons to the pairing-mediating fluctuations, in analogy to strongly-coupled phonon-mediated superconductors, where an enhancement of the Pauli limit was found to reflect the renormalization of the density of states at the Fermi level~\cite{bulaevskii1988,schossmann1989}. The maximum of $H_{\mathrm{c2}}^{\perp}$ near $P_{0}$ in \CRA\ resembles the case of other Ce-based heavy-fermion superconductors, such as CeIrSi$_{3}$~\cite{settai2008} and CeCu$_2$Si$_2$~\cite{lengyel2011}, where $H_{\mathrm{c2}}$ peaks close to the critical pressure of the magnetic order. The atypical shape of $H_{\mathrm{c2}}^{\perp}(T)$ can then be explained by a sizable increase of the coupling strength upon approaching the QCP with decreasing temperature.

\begin{figure}[!ht]
	\includegraphics[width=\columnwidth]{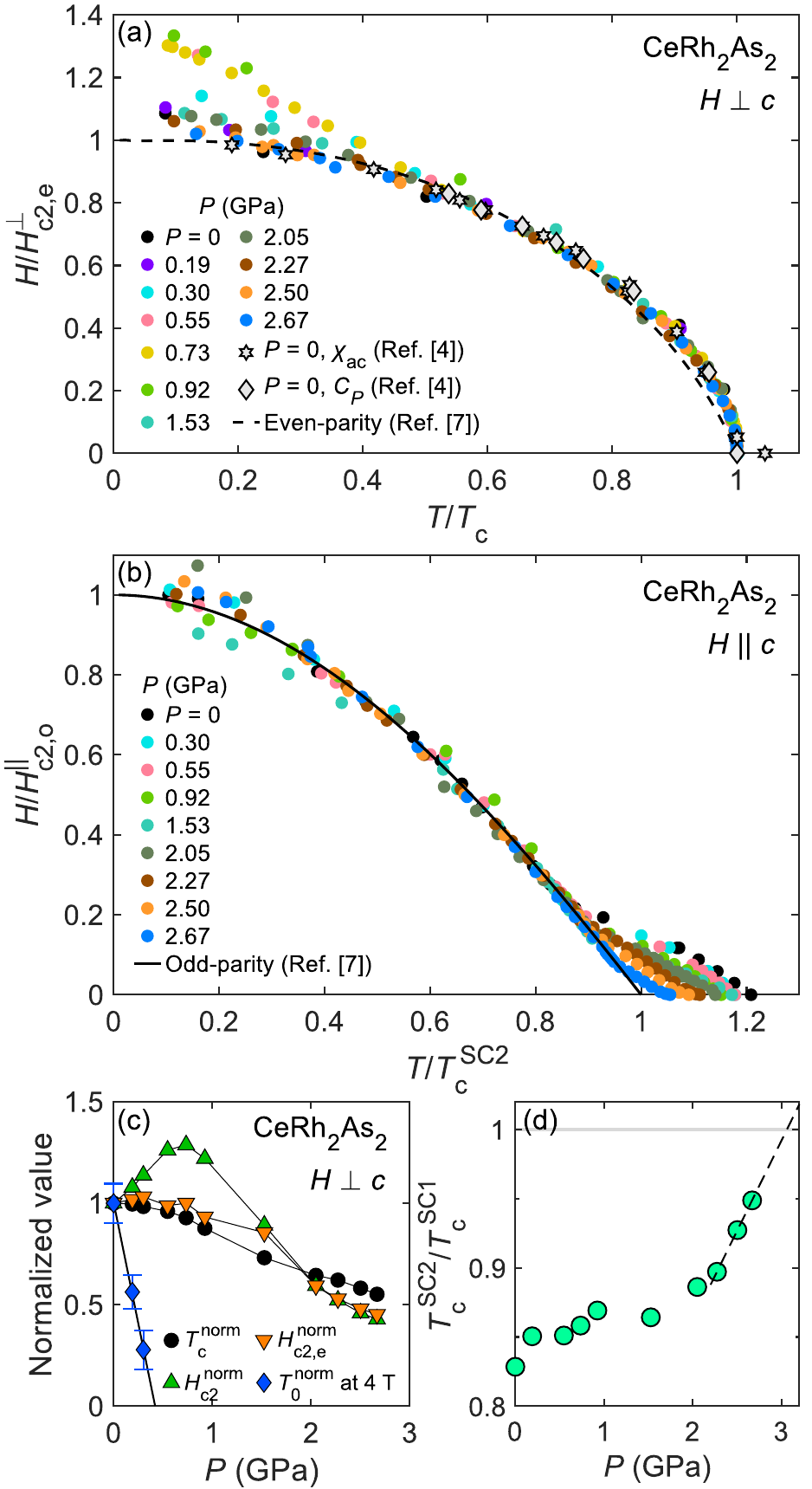}
	\caption{Summarized effect of pressure ($P$) on the superconductivity in \CRA. (a) Scaled in-plane critical field curves (\Tc -- zero-field critical temperature; $H_{\textrm{c2,e}}^{\perp}$ -- effective Pauli-limiting field, based on a theoretical even-parity curve~\cite{khim2021,landaeta2022}, shown as the dashed line). Scaled critical field data from the earlier heat capacity ($C_{\textrm{P}}$) and AC magnetic susceptibility ($\chi_{\textrm{AC}}$) measurements~\cite{khim2021} are included. (b) Scaled out-of-plane critical field curves (\Tcodd -- extrapolated zero-field critical temperature of the high-field SC2 phase; $H_{\textrm{c2,o}}^{||}$ -- critical field at $T=0$ according to a fit to the odd-parity model~\cite{khim2021,landaeta2022}, shown as the solid line). (c) Pressure dependence of the relevant properties, normalized to their zero-pressure values, including the critical temperature \To\ of Phase~I at 4\,T~\cite{pfeiffer2024} ($H_{\textrm{c2}}^{\perp}$ -- the actual in-plane upper critical field extrapolated to $T=0$). (d) Pressure dependence of the ratio of the SC2 and SC1 critical temperatures at $H=0$.}
\label{fig:Discussion}
\end{figure}

By multiplicatively scaling the $H_{\mathrm{c2}}^{\perp}(T)$ data in Fig.~\ref{fig:PhaseDiagrams_ab} along the temperature and field axes, the critical field curves at different pressures can be shown to follow the same form down to at least $T/T_{\mathrm{c}}\approx0.5$. This also applies to the ambient-pressure critical field data from the previous heat capacity and magnetic AC susceptibility measurements~\cite{khim2021,landaeta2022}. Note, that in the case of heat capacity and susceptibility, the field has to be scaled by a factor of 2, as opposed to 2.5 for the ambient-pressure resistivity, which reflects the difference in \Tc\ (addressed in SM). This universal behavior can be decently approximated by the theoretical even-parity curve from an earlier study~\cite{khim2021} [the dashed curve in Fig.~\ref{fig:Discussion}(a)], with the agreement even extending to the lowest achieved temperatures at 2.50 and 2.67\,GPa. Using the $T/T_{\mathrm{c}}>0.5$ portion of the $H_{\mathrm{c2}}^{\perp}(T)$ data to constrain the theoretical even-parity curve, we define the effective zero-temperature Pauli limit $H_{\mathrm{c2,e}}^{\perp}$, which serves as the scaling factor in Fig.~\ref{fig:Discussion}(a) and does not account for the low-temperature deviation below 2\,GPa. The decrease of this Pauli limiting field with pressure can mostly be correlated with the reduction of \Tc\ [Fig.~\ref{fig:Discussion}(c)].

Close to \Tc, the theoretical curve visibly deviates from the data, implying that the initial slope of $H^{\perp}_{\mathrm{c2}}(T)$ and hence the in-plane orbital limiting field estimated from it are significantly steeper than found in the earlier experiments~\cite{khim2021,landaeta2022}, where the slope was possibly underestimated due to a low density of points. Based on the current data, we place the lower limit of $|(dH^{\perp}_{\mathrm{c2}}/dT)_{T_{\mathrm{c}}}|$ at 113\,T/K at $P=0$. A steep initial slope is indeed expected for superconductivity arising from electronic states with extreme effective mass anisotropy or belonging to quasi-two-dimensional elements of the Fermi surface~\cite{hafner2022,wu2024,chen2024}. The quasi-two-dimensional Fermi sheets could also explain why the model based on loosely coupled superconducting layers~\cite{yoshida2012} may work for bulk \CRA.

Aside from the influence of the quantum critical fluctuations, the atypical shape of $H_{\mathrm{c2}}^{\perp}$ at low temperature could also be caused by a presence of a finite-momentum pairing state~\cite{fulde1964,larkin1965}. A particular variant of it, called a complex stripe phase, has been predicted for systems with locally broken inversion symmetry~\cite{yoshida2013}. The superconductivity in \CRA\ is strongly Pauli limited for $H\perp c$ and the band structure is likely anisotropic~\cite{hafner2022,cavanagh2022,wu2024,chen2024} --- these features are necessary for enabling a finite-momentum pairing~\cite{matsuda2007}. However, such states are easily destroyed by impurities~\cite{matsuda2007}. We estimate comparable mean free paths $l$ and coherence lengths $\xi$ for the probed batch of \CRA\ samples (see SM). For the in-plane values we find $l_{\mathrm{ab}} \approx 2 \xi_{\mathrm{ab}}$, meaning that the clean limit is possibly applicable for $H \parallel c$. Further studies using bulk probes and newer generations of crystals are needed for investigating the anomalous shape of $H_{\mathrm{c2}}^{\perp}$ and the possible existence of the complex stripe phase.

As for the effect of pressure on the multi-phase superconductivity, both SC1 and SC2 persist well beyond the Phase~I critical pressure of $P_0 \sim0.5$\,GPa~\cite{pfeiffer2024}. Besides reinforcing the fact that the SC1-SC2 line is not a transition between a mixed SC+I and a pure SC state~\cite{semeniuk2023}, this finding also bears consequences for the spin-flop interpretation of the $H^{*}$ signature, suggested in previous works~\cite{machida2022,ogata2023}. Namely, if the SC1-SC2 transition involves magnetic ordering, then the latter is entirely separate from Phase~I, and is at least as robust under pressure as $H^{*}$. Hence, while the \To\ quantum critical transition is a likely source of pairing-mediating interactions, the phenomenology of the SC1-SC2 transition can be regarded independently of Phase~I.

The $H_{\mathrm{c2}}^{||}(T)$ curve of the SC2 state retains the orbitally limited character at all pressures. As shown in Fig.~\ref{fig:Discussion}(b), by multiplicatively scaling the data for different pressures along the temperature and field axes, the phase boundaries between the SC2 and normal states can be superimposed onto a single theoretical curve for an odd-parity state~\cite{landaeta2022}. Such a universality indicates that changes to the effective quasiparticle mass and \Tc\ are sufficient for explaining the pressure dependence of the SC2 critical field curve. Pressure dependence of the $c$-axis orbital limiting field indicates that the Fermi velocity increases above 1.5\,GPa (see SM), which possibly reflects the weakening of correlations at high pressure observed in the normal-state resistivity~\cite{pfeiffer2024}.

By extrapolating the SC2 critical field curve to $H=0$, we define a fictive critical temperature \Tcodd\ and denote the usual \Tc\ as \Tceven. The ratio \Tcodd$/$\Tceven\ increases with pressure and steeply approaches unity for ${P>2}$\,GPa [Fig.~\ref{fig:Discussion}(d)], in agreement with $H^{*}$ decreasing towards zero [Fig.~\ref{fig:PhaseDiagrams_c}(c)]. 

Theories describing locally non-centrosymmetric superconductors quantify the extent of local inversion asymmetry with a parameter $\alpha/t_{\mathrm{c}}$, where $\alpha$ is the strength of Rashba interaction and $t_{\mathrm{c}}$ is the interlayer hopping parameter. According to the parity-switching picture, the observed enhancement of \Tcodd\ relative to \Tceven\ implies an increase of $\alpha/t_{\mathrm{c}}$ under pressure, whereas one would naively expect $t_{\mathrm{c}}$ to become larger as the $c$-axis lattice parameter shrinks, and thus $\alpha/t_{\mathrm{c}}$ to decrease with pressure. The increase of $\alpha/t_{\mathrm{c}}$ could be linked to the prediction of the Fermi sheets relevant for the superconductivity being centered at the edges of the Brillouin zone (BZ), where the non-symmorphic symmetry of the lattice enforces $t_{\mathrm{c}}=0$ \cite{cavanagh2022,hafner2022}. If these Fermi sheets shift closer to the BZ boundaries under pressure, the mean value of $t_{\mathrm{c}}$ can indeed decrease. Note that the increase of $\alpha/t_{\mathrm{c}}$ should lead to a further enhancement of the Pauli limiting field of SC1. For $H\parallel c$, the effect should be particularly strong, but the SC1 section of the phase diagram becomes too small to investigate that. For $H\perp c$, however, the Pauli limiting field is expected to increase only by a minor amount (discussed in SM) in line with our results.

What is more striking, is the unexpectedly steep increase of \Tcodd$/$\Tceven\ above 2\,GPa. The trend predicts that above 3\,GPa the SC2 phase will become the only stable SC phase, in agreement with the pressure dependence of $H^*$. According to the model, when the staggered Rashba interaction is present, the ratio of the magnitudes of the odd- and even-parity state order parameters is given by $(\nicefrac{\alpha}{t_{\mathrm{c}}})/\sqrt{(\nicefrac{\alpha}{t_{\mathrm{c}}})^2+1}$, implying that \Tcodd\ is always smaller than \Tceven. This inference, however, assumes the equal bare pairing potentials of the two states (before the Rashba interaction is considered)~\cite{cavanagh2022,fischer2023}. The strong changes of \Tcodd$/$\Tceven\ (and $H^{*}$) at high pressure challenge this assumption and provide a new input for extended theories.

\textit{Conclusions.}---Our study of the superconductivity in \CRA\ under pressure demonstrates a strong overall sensitivity of SC properties to the proximity to the quantum critical point. However, the pronounced stabilization of the SC2 state over SC1 under pressure indicates that a different tuning parameter drives the balance between the two SC phases. The proposed relevance of the staggered Rashba interaction fits this picture and implies a major increase of the local non-centrosymmetry with pressure. These results demonstrate a good tunability of the system and motivate further experimental and theoretical investigations into the microscopic processes involved in the superconductivity and other ordered states of \CRA.

\textit{Acknowledgments.}---We thank D. Agterberg, K. Ishida, S. Kitagawa, Y. Yanase, G. Knebel, D. Aoki, J.-P. Brison, D. Braithwaite, A. Ramires, G. Zwicknagl, C. Timm, J. Link, and M. Grosche for fruitful discussions. We acknowledge funding from  Deutsche Forschungsgemeinschaft (DFG) for the CRC 1143 - Project No. 247310070 (project C10) and for the Wuerzburg-Dresden cluster of excellence EXC 2147 ct.qmat Complexity and Topology in Quantum Matter - Project No. 390858490.


\begin{center}
\textbf{\large Supplemental Material}
\end{center}

\setcounter{equation}{0}
\setcounter{figure}{0}
\setcounter{table}{0}
\makeatletter
\renewcommand{\theequation}{S\arabic{equation}}
\renewcommand{\thefigure}{S\arabic{figure}}
\renewcommand{\thetable}{S\arabic{table}}

\section{Methods}

Data presented in the current letter were obtained throughout the same set of high-pressure experiments as our previous work on \CRA~\cite{pfeiffer2024}. Many of the details pertaining to the sample preparation, measurement techniques, and experimental protocol are the same for both works. We do not reproduce all the information here and instead refer the reader to the Supplemental Material of our previous paper~\cite{pfeiffer2024}. Details that are specifically relevant to the present study of the superconducting (SC) properties of \CRA\ under hydrostatic pressure are provided below.

\subsection{Samples of C\lowercase{e}R\lowercase{h}\textsubscript{2}A\lowercase{s}\textsubscript{2}}

The pressure cell was initially closed with three samples of \CRA, denoted as Sample~1, Sample~2, and Sample~3. These belonged to the same batch of crystals that was used for the early studies of the material in Refs.~\cite{khim2021,hafner2022,landaeta2022,kibune2022,ogata2023}. These samples had the ratio of resistivity at 300\,K to the one at 0.5\,K equal to 2.0, and their zero-temperature $C/T$ value was $\gamma=1.2$\,J\,K$^{-2}$\,mol$^{-1}$. A previously measured resistivity value of 125\,$\micro\Omega$\,cm at 300\,K was used as a reference for converting measured resistance to resistivity. Sample~1 and Sample~2 were cut from the same single crystal and had their $c$ axes oriented along the applied magnetic field. The initial zero-pressure run and the 0.30\,GPa run are exceptions: Sample~2 was initially mounted with the field perpendicular to the $c$ axis, which was changed after the 0.30\,GPa run, during which Sample~1 was found to be misaligned. Sample~3 was always oriented with the $c$ axis perpendicular to the field. Resistivity data presented in our previous work~\cite{pfeiffer2024} come from the measurements on Sample~3.

\subsection{Sequence of applied pressures}

At the start of the experiment, the pressure cell was sealed with a small force (<2\,kN) and the first round of measurements was conducted with the samples at effectively ambient pressure. Afterwards, pressure (in GPa) was applied in the following order: 0.30, 0.19, 0.55, 0.92, 1.53, 0.73, 2.05, 2.50, 2.67, 2.27. This sequence was interrupted after 0.30, 0.55, and 1.53\,GPa, when the cell had to be opened in order to repair contacts of Sample~1 or fix its alignment. After 1.53\,GPa, Sample~1 could not be recovered and was replaced with Sample~4, coming from a newer batch of \CRA, also used in Ref.~\cite{semeniuk2023}. Whenever the cell was opened and closed again, a control measurement at effectively ambient pressure was conducted first, in order to ensure that the orientation of the samples was not disturbed during the closure. Since the SC2 state is very sensitive to the magnetic field direction, the fact that we could clearly detect SC2 at high pressures demonstrates that the alignment of the samples was not substantially affected by the application of pressure.

\subsection{Measurements at low temperatures and high magnetic fields}

\begin{figure}[t]
        \includegraphics[width=\linewidth]{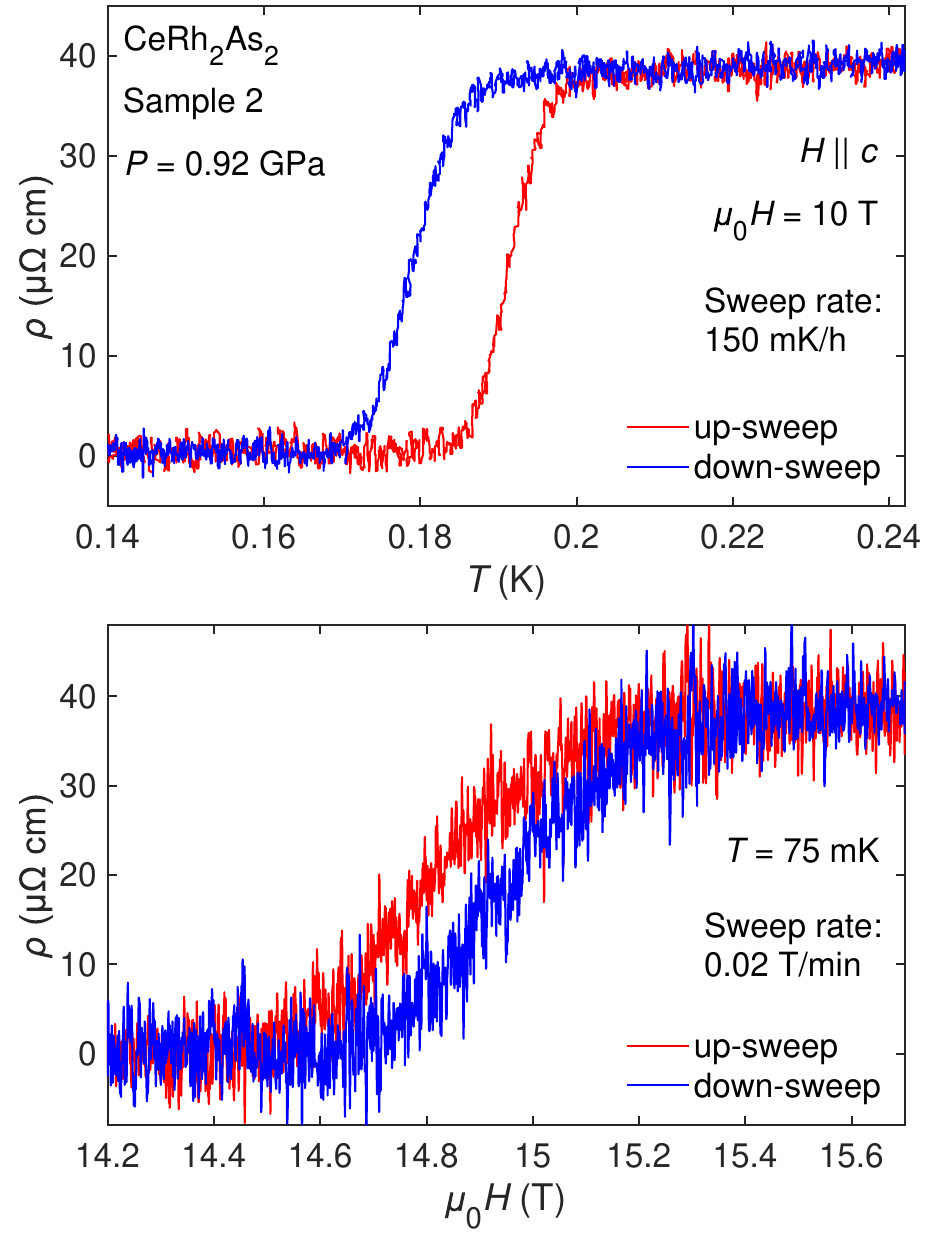}
    \caption{Examples of irreversibility of the signal in temperature sweeps due to a thermal lag between the pressure cell and the thermometer (top) and in magnetic field sweeps due to a magnetocaloric effect (bottom).}
        \label{sfig:lag}
\end{figure}

In order to tackle the thermal lag caused by the significant increase of the heat capacity of the beryllium copper pressure cell at low temperatures~\cite{karaki1997}, and by its even stronger enhancement and high magnetic fields, we typically conducted temperature sweeps in both directions, keeping the rate between 200 and 50\,mK/h. When determining \Tc, the effect of the lag was eliminated by taking an average between the two \Tc\ values, extracted from the sweeps of opposite directions. At fixed magnetic fields of approximately 8\,T and above, achieving stable temperature below about 200\,mK required significantly longer waiting times, reaching more than an hour in the most extreme cases, making temperatures sweeps impractical.
In such regions of the phase space, only magnetic field sweeps were conducted, and care was taken to ensure that the pressure cell was at a stable temperature throughout each sweep. The magnetic field sweeps were conducted at a rate of 0.01-0.02\,T/min, in order to limit the inductive heating and the magnetocaloric effect. Even with this rate, we could observe a shift between the measured SC critical fields in up-/down-sweeps, due to a magnetocaloric heating/cooling. Examples of the irreversible behavior in $\rho(T)$ and $\rho(H)$ are demonstrated in Fig.~\ref{sfig:lag}. The critical field data in the phase diagrams is the average of the two values for sweeps in opposite directions.

We defined the SC critical fields and temperatures as the points of maximum slope in the resistivity drop. The derivatives $d\rho/dT$ and $d\rho/dH$ were calculated using a Savitzky-Golay filter with a second order polynomial. Defining \Tc\ and $H_{\mathrm{c2}}$ as a 50\% threshold of resistivity drop changes the values of \Tc\ and $H_{\mathrm{c2}}$ at most by 5\,mK and 50\,mT, respectively, and does not affect the interpretation of our results.

\section{Comparison of different probing techniques}

\begin{figure}[t]
        \includegraphics[width=\linewidth]{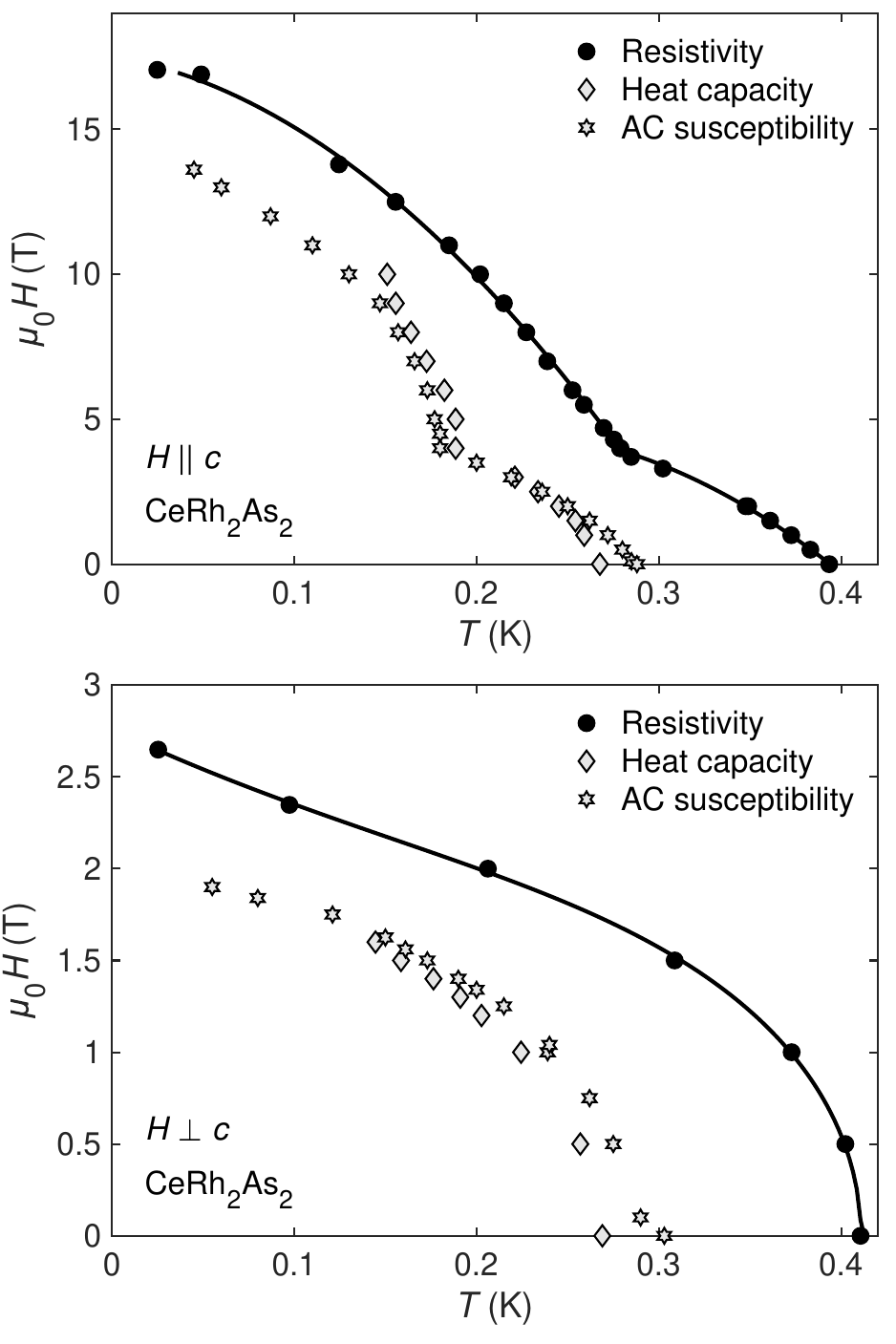}
	    \caption{Comparison of ambient pressure upper critical field curves of \CRA\ obtained using different probes.}
        \label{sfig:probes_comparison}
\end{figure}

A recurring observation across multiple studies of \CRA~\cite{khim2021,hafner2022,semeniuk2023} is the significant difference between the resistive \Tc\ ($\sim0.4$\,K at ambient pressure) and the values obtained from the bulk probes (up to 0.35\,K). In Fig.~\ref{sfig:probes_comparison} we compare the critical field curves of \CRA\ obtained from measurements of different properties: heat capacity~\cite{khim2021}, AC magnetic susceptibility~\cite{landaeta2022}, and resistivity. The same batch of crystals was used for all the measurements. Resistivity displays a significantly larger \Tc\ than seen in the other probes, but the increase of \Tc\ is accompanied by a nearly proportional increase in the critical field, resulting in a similar shape of the phase boundary. For unconventional superconductors, \Tc\ is generally very sensitive to a crystalline disorder. In inhomogeneous samples, upon cooling, a zero-resistance path between the voltage probes can form even when only a minor portion of the sample is superconducting, and it is therefore expected for charge transport probes to report a higher \Tc\ than thermodynamic and magnetic probes. For $H\parallel c$, the SC phase switching occurs at 4\,T, regardless of the probing technique. An improvement of the sample quality causes a significant increase in bulk \Tc\ as detected by the specific heat, while resistive \Tc\ remains rather stable~\cite{semeniuk2023}. Given the aforementioned information, we find it reasonable to attribute the discrepancy between \Tc\ in the thermodynamic and charge transport measurements to the resistivity signal being dominated by parts of the sample where \Tc\ is locally enhanced due to a more optimal stoichiometry, reduced impurity content, or favorable defect-induced strain.

\section{Resistivity near the superconducting transition as a function of temperature and magnetic field}

\begin{figure*}[t]
        \includegraphics[width=\linewidth]{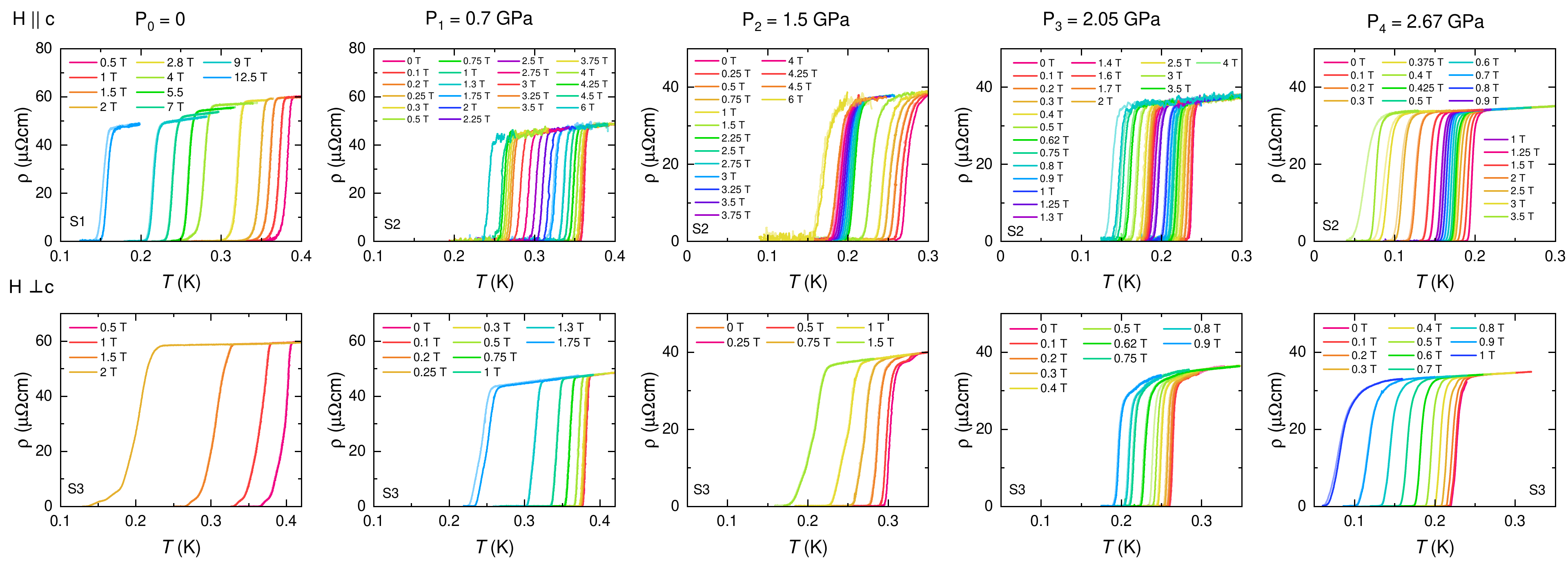}
	    \caption{Resistivity of \CRA\ as a function of temperature at different pressures. The top/bottom rows correspond to the $c$-axis/$ab$-plane field orientation. The labels "S1", "S2", and "S3" indicate Sample~1, Sample~2, and Sample~3, in accordance with Fig.~1. An upward/downward sweep for each given field is plotted in a more/less saturated color.}
        \label{sfig:TSweeps}
\end{figure*}

\begin{figure*}[t]
        \includegraphics[width=\linewidth]{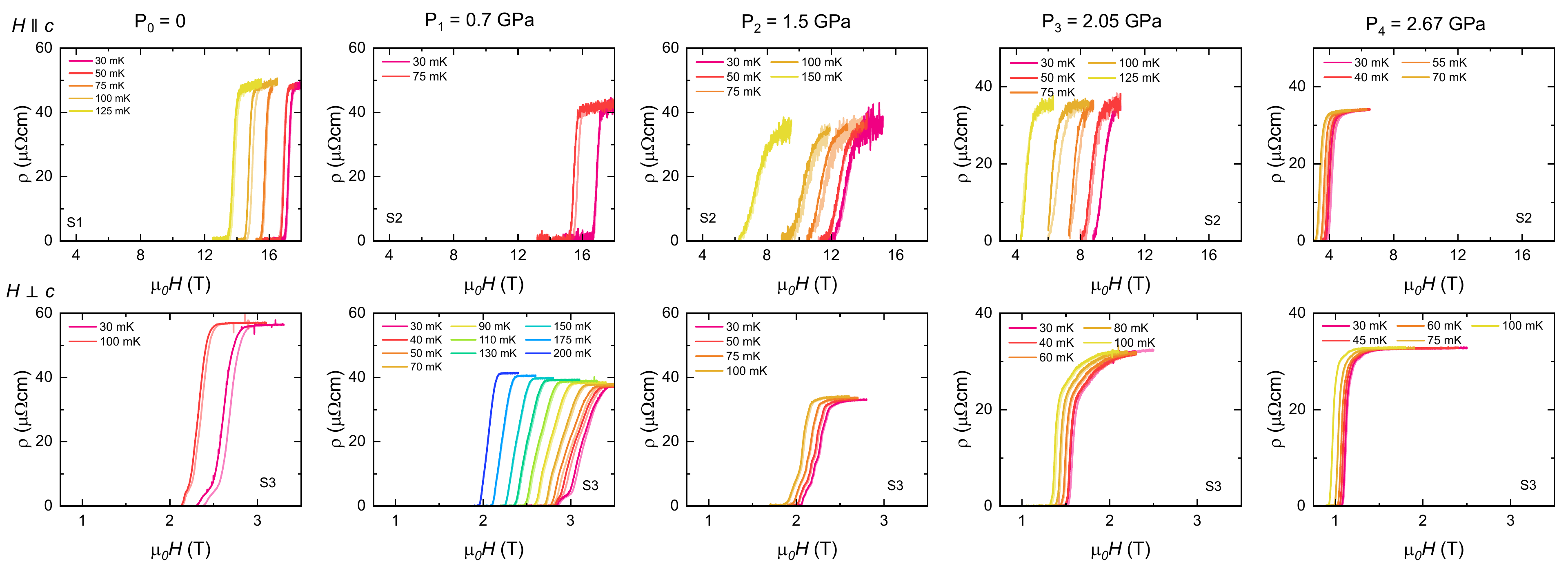}
	    \caption{Resistivity of \CRA\ as a function of magnetic field at different pressures. The top/bottom rows correspond to the $c$-axis/$ab$-plane field orientation. The labels "S1", "S2", and "S3" indicate Sample~1, Sample~2, and Sample~3, in accordance with Fig.~1. An upward/downward sweep for each given temperature is plotted in a more/less saturated color.}
        \label{sfig:HSweeps}
\end{figure*}

In Fig.~\ref{sfig:TSweeps} and Fig.~\ref{sfig:HSweeps} we show a selection of $\rho(T)$ and $\rho(H)$ data for the two magnetic field directions. A moving average smoothing filter was applied to the data for Sample~2 at 0.7, 1.5, and 2.05\,GPa. In the other plots, the raw data are shown.

\section{Influence of the local non-centrosymmetry on the in-plane critical field}

\begin{figure}[t]
        \includegraphics[width=\linewidth]{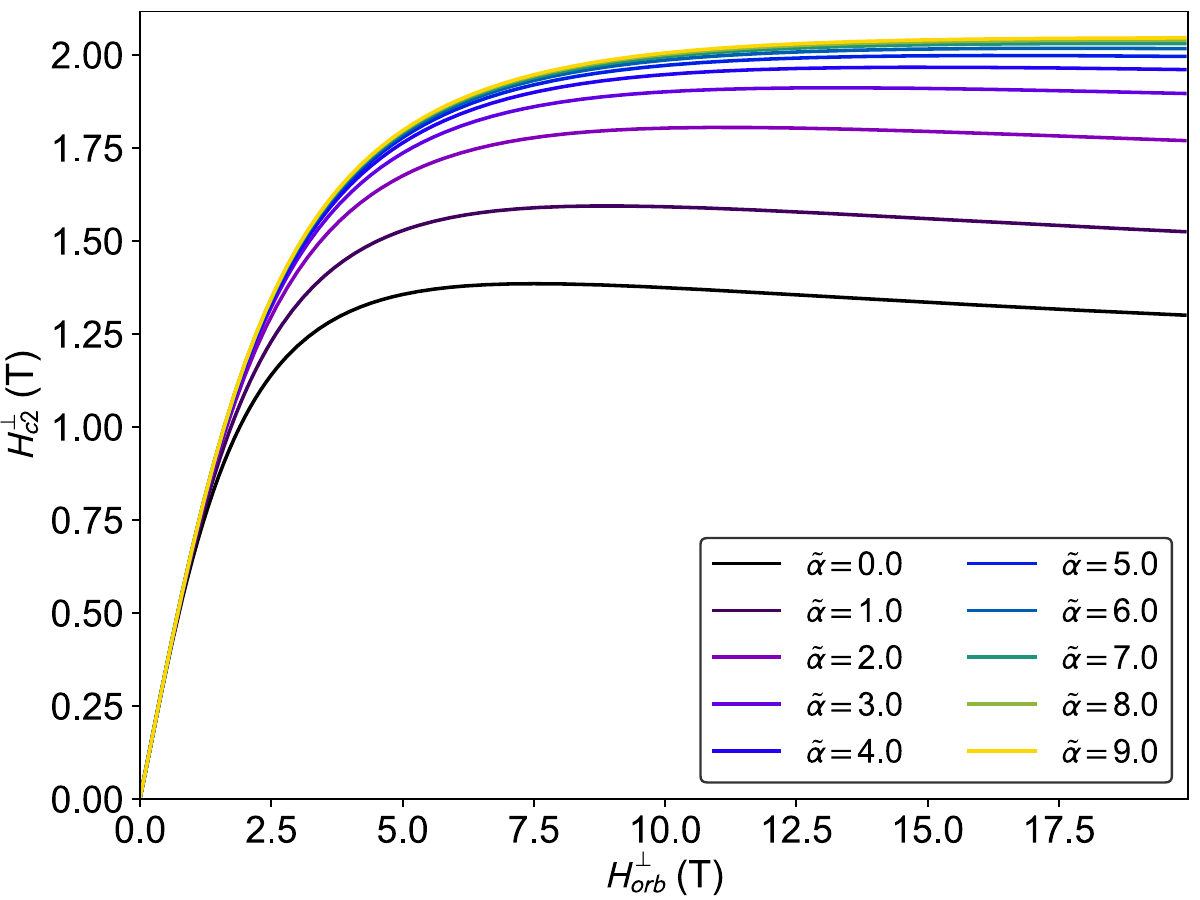}
	    \caption{Calculated in-plane upper critical field $H_{\mathrm{c2}}^{\perp}$ of \CRA\ against the in-plane orbital limiting field $H_{\mathrm{orb}}^{\perp}$ for different values of the the local non-centrosymmetry parameter $\widetilde{\alpha}$. The calculations were performed according to the previously used model~\cite{landaeta2022}, assuming the even-parity superconducting state. The calculations used the temperature value of 10\,mK.}
        \label{sfig:effect_of_Rashba_on_Hc2ab}
\end{figure}

In this paragraph we show why a sizable increase of $\alpha/t_{\mathrm{c}}$ with pressure, deduced from the behavior of $T_{\mathrm{c}}^{\mathrm{SC2}}/T_{\mathrm{c}}^\mathrm{SC1}$ against $P$, is not expected to significantly enhance the in-plane upper critical field of the SC1 state. Using the even-/odd-parity model~\cite{landaeta2022} it is possible to evaluate the effect of local non-centrosymmetry parameter $\widetilde{\alpha}$ on the low-temperature in-plane upper critical field $H_{\mathrm{c2}}^{\perp}$ of \CRA. The resultant value is displayed in Fig.~\ref{sfig:effect_of_Rashba_on_Hc2ab} for different values of $\alpha/t_{\mathrm{c}}$, as a function of the orbital limiting field. The previously found bare Pauli limit of 1.47\,T was used for the calculations~\cite{khim2021}. The zero-temperature limit was approximated by using a temperature value of 10\,mK in the calculations. When the orbital limit exceeds 5\,T (a condition that is fulfilled for all pressures for $H \perp c$, as is demonstrated in the next section), it no longer has a significant effect on the critical field, as the superconductivity becomes strongly Pauli limited. An increase of $\widetilde{\alpha}$ leads to a substantial enhancement of the upper critical field up to an asymptotic limit. This effect is most pronounced for values up to about 4, at which point the critical field is already within less than 10\% of the limit. The actual effective value of $\widetilde{\alpha}$ for \CRA\ at zero pressure was estimated to be 3.4~\cite{khim2021}, meaning that further increase of $\widetilde{\alpha}$ is expected to result in no more than approximately 10\% enhancement of the in-plane critical field.

\section{Coherence length and mean free path estimation}

In this section we investigate whether the clean limit of superconductivity is achieved in our samples of \CRA\ by comparing estimated mean free paths and coherence lengths for different crystallographic directions.

We estimate the mean free paths in \CRA\ by referring to the Wiedemann-Franz law, which appears to be upheld reasonably well at low temperature, based on the earlier measurements of thermal conductivity $\kappa$~\cite{onishi2022}. Given the normal state low-temperature resistivity $\rho$, the law states that:

\begin{equation}
    \label{eq:wf}
	L = \frac{\rho \kappa}{T}
\end{equation}

\noindent
where $L=2.44\times10^{-8}$\,V$^{2}$\,K$^{-2}$ is the Lorenz constant. Within the Drude model, electronic thermal conductivity can be expressed as:

\begin{equation}
    \label{eq:drudett}
	\kappa = C_{v} v_{\mathrm{F}} l/3
\end{equation}

\noindent
where $C_v$ is the electronic heat capacity per unit volume, $v_\mathrm{F}$ is the Fermi velocity, and $l$ is the mean free path. Combining this expression with the Wiedemann-Franz law gives:

\begin{equation}
    \label{eq:mfp1}
    l = \frac{3 L T}{C_{\mathrm{v}} \rho v_{\mathrm{F}}} = \frac{3 L}{\gamma_{\mathrm{v}} \rho v_{\mathrm{F}}}
\end{equation}

\noindent
where we introduced the volumetric Sommerfeld coefficient $\gamma_{\mathrm{v}}=C_{\mathrm{v}}/T$.

Next, we use the BCS formula relating the coherence length $\xi$ to the Fermi velocity via the magnitude of the SC gap function $\Delta$. We admit that it is not rigorous to use BCS expressions in the case of \CRA, and the following results should be regarded with caution. The formula states:

\begin{equation}
    \xi = \frac{\hbar v_{\mathrm{F}}}{\pi \Delta}
\end{equation}

\noindent
where $\hbar$ is the reduced Planck constant. We then use the phenomenological Ginzburg-Landau coherence length extracted from the orbital limiting field $H_{\mathrm{orb}}$, as a proxy for the microscopic BCS coherence length. Keeping in mind that the orbital motion occurs in the plane perpendicular to the field direction:

\begin{equation}
\label{eq:coh_len_ab1}
\mu_{0}H_\mathrm{orb}^{||} = \frac{\Phi_{0}}{2\pi\xi_{ab}^{2}}
\end{equation}

\begin{equation}
\label{eq:coh_len_c1}
\mu_{0}H_\mathrm{orb}^{\perp} = \frac{\Phi_{0}}{2\pi\xi_{ab}\xi_{c}}
\end{equation}

\noindent
where $\Phi_{0}$ is the flux quantum, and $\mu_{0}$ is the permeability of free space. We use the bulk orbital limit of ${\mu_{0}H_{\mathrm{orb}}^{||}=17.7}$\,T obtained in the earlier study from the slope of $H_{\mathrm{c2}}^{\parallel}(T)$ close to \Tc~\cite{khim2021}. As for ${\mu_{0}H_{\mathrm{orb}}^{\perp}}$, we note the visible mismatch between the experimental and modelled data near \Tc\ in Fig.~4(b), and propose that the previously found in-plane orbital limit of 7.4\,T was a sizable underestimate. In the next paragraph, we describe how a new estimate was obtained.

\begin{figure}[t]
        \includegraphics[width=\linewidth]{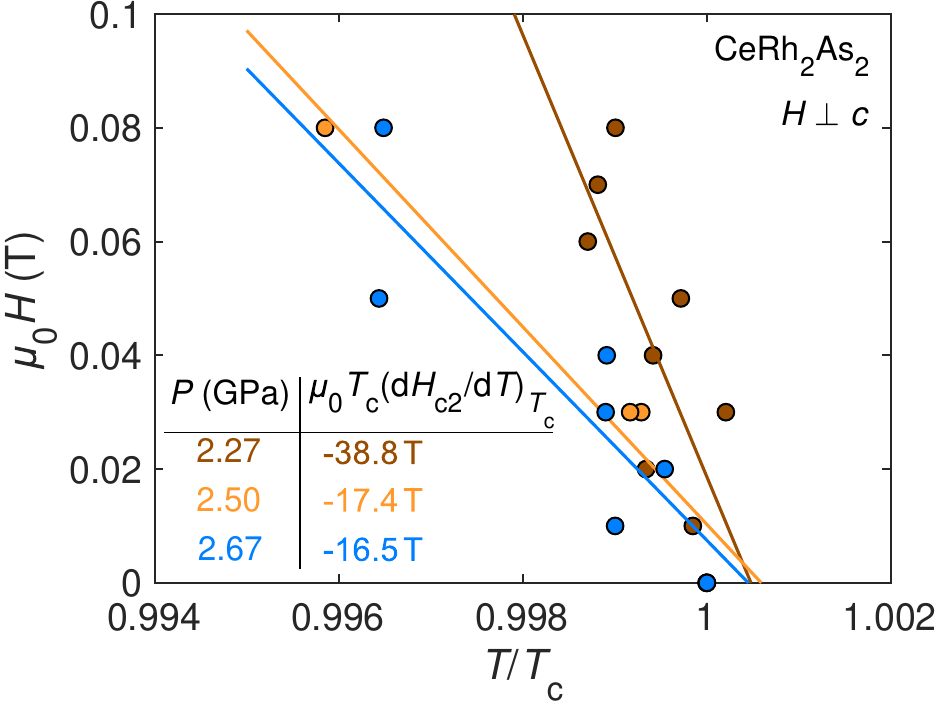}
	    \caption{Estimation of the initial slope of the critical field curve for field perpendicular to $c$ axis, at the three highest pressures. The slope was determined by fitting a line to the data in the 0 to 0.1 T field range.}
        \label{sfig:Hc2_slope_at_Tc}
\end{figure}

The critical field curve for $H\perp c$ becomes very steep close to \Tc, and the random uncertainty in temperature of $\approx 0.5$\,mK translates to a sizable uncertainty in the slope $dH_{\mathrm{c2}}/dT|_{T_{\mathrm{c}}}$. We illustrate it in Fig.~\ref{sfig:Hc2_slope_at_Tc}, for the three highest pressures of the study, for which the critical field curve close to \Tc\ was sampled at smaller intervals compared to the other pressures. Based on the examined data, we place a conservative upper limit on $\mu_{0}T_{\mathrm{c}}(dH_{\mathrm{c2}}/dT)_{T_{\mathrm{c}}}$ at -17\,T at 2.5\,GPa, taking $T_{\mathrm{c}}=0.3$\,K (the bulk value). If we then assume that the scaling depicted in Fig.~4(a) correctly captures changes to the critical field curve very close to \Tc, the slope should have double the value at ambient pressure: -34\,T. This gives the orbital limit of at least 24\,T. Using the relationship $H_{\mathrm{orb}}=\eta T_{\mathrm{c}}|dH_{\mathrm{c2}}/dT|_{T_{\mathrm{c}}}$ given by the Werthamer-Helfand-Hohenberg theory~\cite{helfand1966}, this corresponds to the orbital limit of at least 24\,T ($\eta$ is equal to 0.69 for the dirty limit and to 0.73 for the clean limit).

The resultant coherence lengths are:

\begin{equation}
\label{eq:coh_len_ab2}
\xi_{ab} = \sqrt{\frac{\Phi_{0}}{2\pi\mu_{0}H_\mathrm{orb}^{||}}}=3.8\,\textrm{nm}
\end{equation}

\begin{equation}
\label{eq:coh_len_c2}
\xi_{c} = \frac{\Phi_{0}}{2\pi\xi_{ab}\mu_{0}H_\mathrm{orb}^{\perp}}=3.6\,\textrm{nm}
\end{equation}

We then apply the BCS weak-coupling approximation ${\Delta=1.76k_{\mathrm{B}}T_{\mathrm{c}}}$ ($k_{\mathrm{B}}$ -- Boltzmann constant) and using $T_{\mathrm{c}}=0.3$\,K (the bulk value) we obtain the Fermi velocities:

\begin{equation}
\label{eq:vF_ab}
v_{\mathrm{F}}^{ab} = 1.76 \pi k_{\mathrm{B}} T_{\mathrm{c}} \xi_{ab} / \hbar = 825\,\textrm{m\,s}^{-1}
\end{equation}

\begin{equation}
\label{eq:vF_c}
v_{\mathrm{F}}^{c} = 1.76 \pi k_{\mathrm{B}} T_{\mathrm{c}} \xi_{c} / \hbar = 780\,\textrm{m\,s}^{-1}
\end{equation}

We use ${\gamma_{\mathrm{m}}=1}$\,J\,mol$^{-1}$\,K$^{-2}$ for the molar Sommerfeld coefficient, according to the low-temperature heat capacity~\cite{khim2021}. With the molar mass of \CRA\ of ${M=495.8}$\,g\,mol$^{-1}$ and the density of $\delta=9.11$\,g\,m$^{-3}$, the corresponding volumetric Sommerfeld coefficient is:

\begin{equation}
\label{eq:gamma_vol}
\gamma_{\mathrm{v}} = \gamma_{\mathrm{m}}\delta/M = 18.4\,\textrm{kJ}\,\textrm{m}^{-3}\,\textrm{K}^{-2}
\end{equation}

Close to $T_{\mathrm{c}}$, the anisotropy of resistivity is $\rho_{c}/\rho_{ab}=2.17$~\cite{mishra2022}. Applying this ratio to our samples, $\rho_{ab}=55$\,$\micro\Omega$\,cm then corresponds to $\rho_{c}=119$\,$\micro\Omega$\,cm. Finally, the mean free paths are:

\begin{equation}
    \label{eq:mfp_ab}
    l_{ab} = \frac{3 L}{\gamma_{\mathrm{v}} \rho_{ab} v_{\mathrm{F}}^{ab}} = 8.8\,\textrm{nm}
\end{equation}

\begin{equation}
    \label{eq:mfp_c}
    l_{c} = \frac{3 L}{\gamma_{\mathrm{v}} \rho_{c} v_{\mathrm{F}}^{c}} = 4.2\,\textrm{nm}
\end{equation}

The obtained values of $\xi$ and $l$ for each direction are not drastically different. While this does not justify assuming the clean limit of superconductivity, it is plausible for it to be realized in less disordered parts of our samples, which charge transport probes are most sensitive to, and particularly in the newer generation of samples, for which $\rho_{ab}=37$\,$\micro\Omega$\,cm just above \Tc.

A larger $ab$-plane orbital limit is also more consistent with the idea of a weak coupling between the sublattices~\cite{yoshida2012} as well as a quasi-two-dimensional character of the Fermi surface observed in recent photoemission experiments~\cite{wu2024, chen2024}. That being said, modelling of the electronic band structure of \CRA\ is still a matter of active ongoing research~\cite{ptok2021,hafner2022,ishizuka2023preprint,kapcia2023preprint}. 

If the quasi-two-dimensional picture of the superconductivity is indeed correct, the orbital limiting field is expected to decrease rapidly for small deviations of $H$ from the $ab$ plane. For a more definitive test, one needs to track the critical temperature using a thermodynamic bulk probe in steps of at least 10\,mT, while ensuring a sub-millikelvin temperature resolution and an exceptionally good alignment of the crystallographic axes with respect to the magnetic field.

\section{Fermi velocity as a function of pressure}

\begin{figure}[t]
        \includegraphics[width=\linewidth]{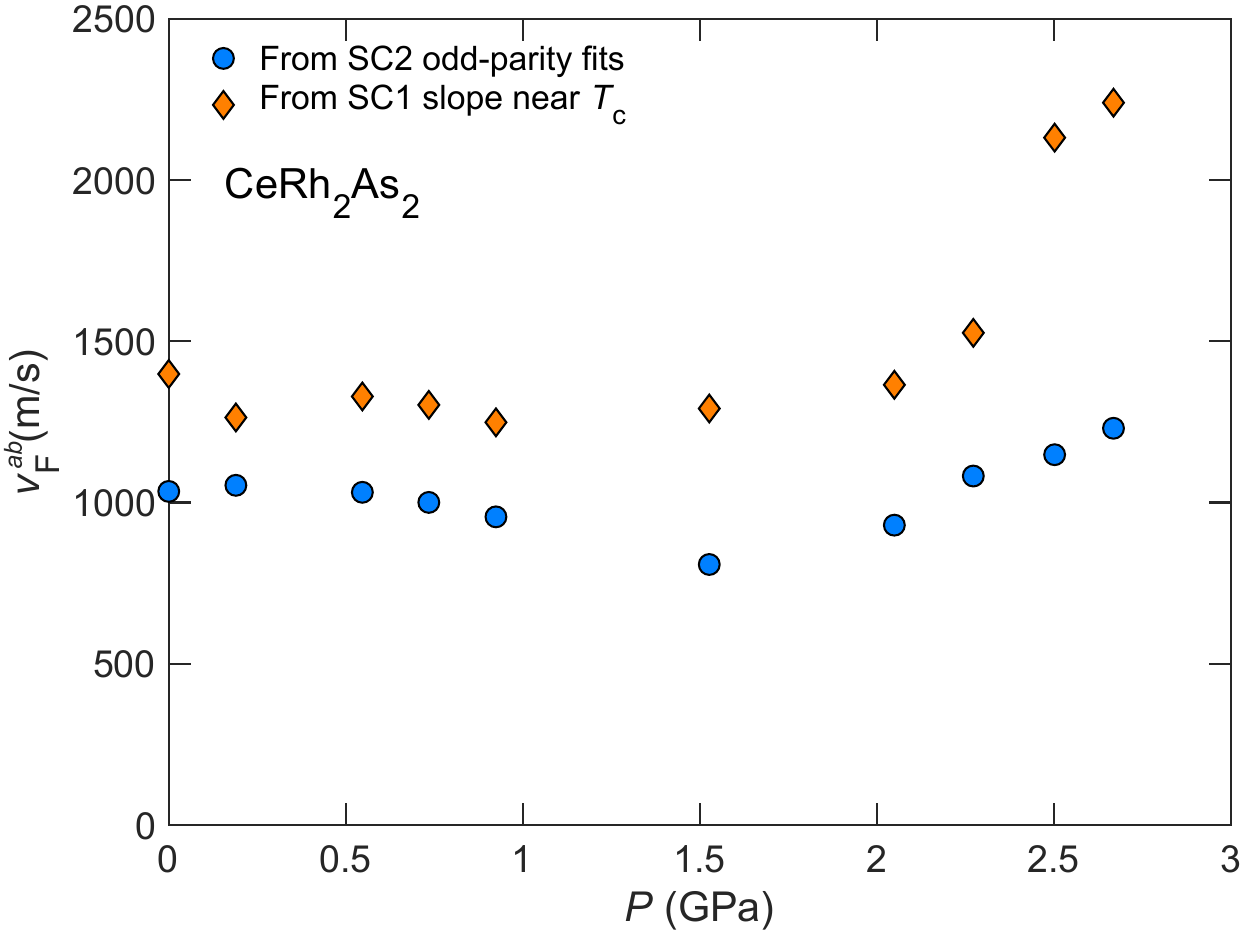}
	    \caption{Fermi velocity in the $ab$ plane as a function of pressure, obtained in two different ways (see the text).}
        \label{sfig:Fermi_velocity}
\end{figure}

Using the pressure dependence of the $c$-axis orbital limiting field, shown in Fig.~2(b), as well as Eq.~\ref{eq:coh_len_ab1} and Eq.~\ref{eq:vF_ab} from the previous section, we examine the change of the $ab$-plane Fermi velocity with pressure. In Fig.~\ref{sfig:Fermi_velocity} we plot the Fermi velocity $v_{\mathrm{F}}^{ab}$ obtained in two ways: taking the orbital limiting field as the zero-temprature limit of the critical field of the SC2 phase (which should be the orbital limit according to the odd-parity interpretation), and estimating the orbital limit from the formula $H_{\mathrm{orb}}^{||}=0.72T_{\mathrm{c}}(dH^{||}_{\mathrm{c2}}/dT)_{T_{\mathrm{c}}}$, where we used the slope of the critical field curve at \Tc~\cite{helfand1966}.

In both cases, we found the Fermi velocity unambiguously increasing at pressures above 1.5\,GPa, which is consistent with the previously observed weakening of correlations at high pressure~[preceding work].

\section{An alternative fit of the even-parity model to the in-plane critical field curves}

\begin{figure}[t]
        \includegraphics[width=\linewidth]{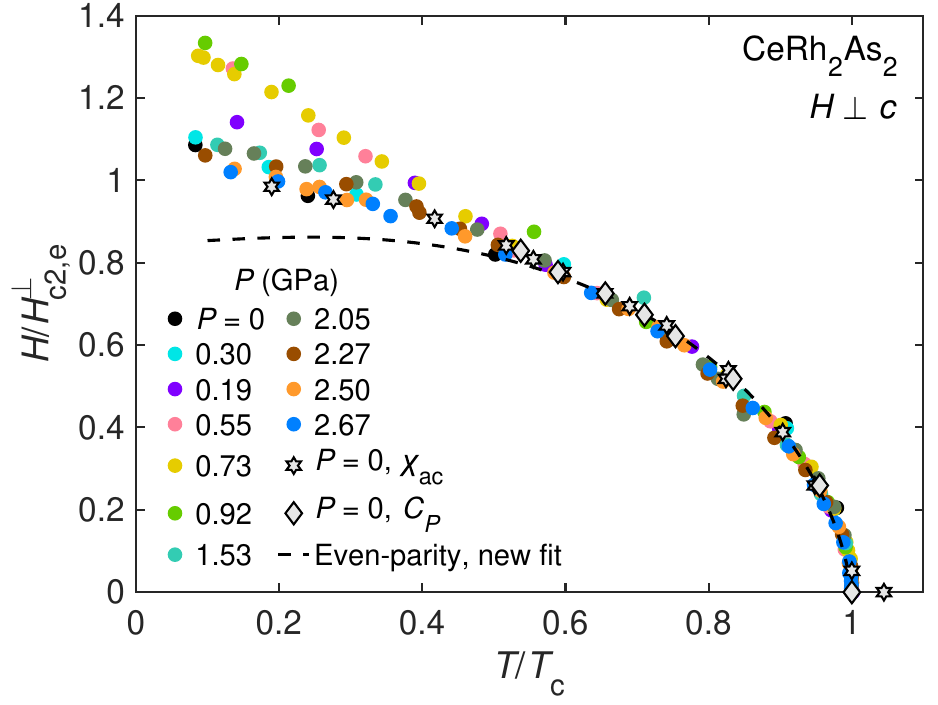}
	    \caption{Plot from Fig.~4(b), with an alternative even-parity model fit, constrained to better match the data near \Tc.}
        \label{sfig:Hc2ab_new_fit}
\end{figure}

Given a noticeable discrepancy between the critical field curves for $H\perp c$ and the even-parity fit close to \Tc\ [Fig.~4(a)], we adjusted the fit, prioritising the region above $0.6T_{\mathrm{c}}$. The result is shown in Fig.~\ref{sfig:Hc2ab_new_fit}. We find that the model can account for the large slope near \Tc\ but cannot simultaneously reproduce the critical field curve for $T\ll T_{\mathrm{c}}$. As discussed in the main text, this could be due to the coupling strength being significantly stronger for $T\ll T_{\mathrm{c}}$ than near \Tc, with the difference most pronounced near 0.7\,GPa, when the system is moved directly towards the quantum critical point upon cooling. Since the model does not account for the temperature dependence of the effective mass, the observed discrepancies are expected, however they become less pronounced at higher pressures.

\section{Addressing the contradictions with S\lowercase{iddiquee et al}., PRB 108, L020504}

In this section, we provide a number of pertinent comments on the recently published independent high-pressure resistivity study of \CRA~\cite{siddiquee2023}. Our intention is to explain the contradictions between some of the results of the present work and the careful statements made in Ref.~\cite{siddiquee2023}.

1). The authors of Ref.~\cite{siddiquee2023} draw attention to the anomaly at 2.5\,GPa in the $T_\mathrm{c}(P)$ curve, hinting at a possible pressure-induced transition. However, the authors stay cautious by stating that the pressure transmitting medium solidifies at room temperature at this exact pressure. They then formulate their conclusion in the abstract carefully: "Above 2.5\,GPa, there might be a second dome of superconductivity, which requires further investigation." The anomaly is absent in our data for three different samples. It is therefore very likely that the discontinuity in $T_{\mathrm{c}}(P)$ at 2.5\,GPa was indeed related to the loss of hydostaticity, which made the applied pressure highly inhomogeneous.

2). Furthermore, in Fig.~3(d) of Ref.~\cite{siddiquee2023}, one can see a pronounced difference between the critical fields in the upward and downward field sweeps. The authors do not argue what values should be the correct ones, analyzing the upward and downwards sweeps independently instead. In the next paragraph, we provide a plausible explanation for the observed difference, and we then discuss the consequences of this ambiguity fruther below.

\begin{figure}[!h]
     \includegraphics[width=\columnwidth]{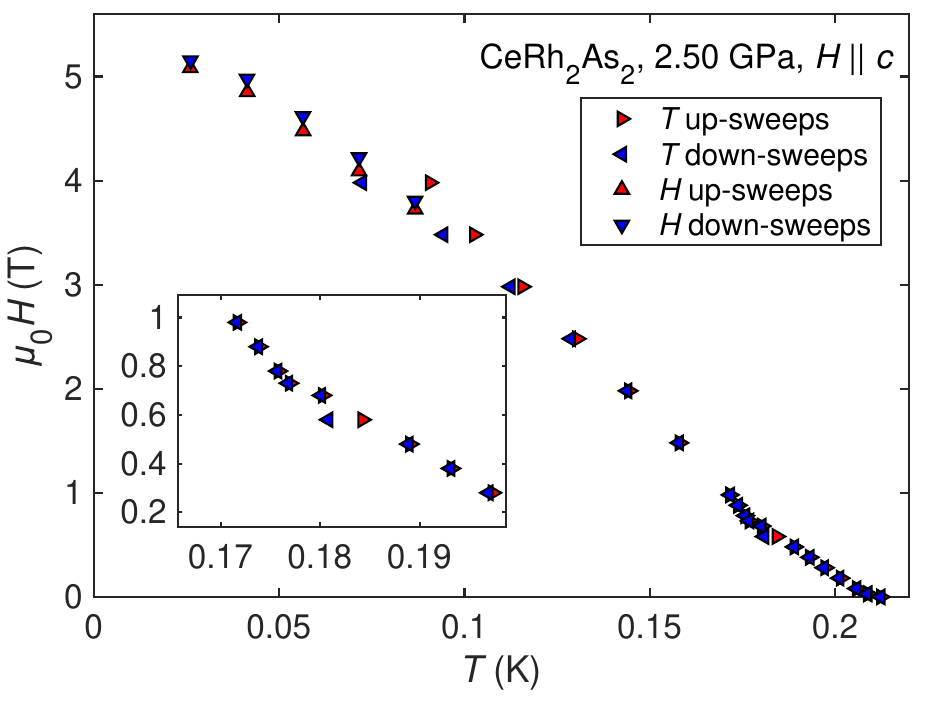}
     \caption{Upper critical field curve of \CRA\ for the $c$-axis magnetic field at 2.50\,GPa. The transition temperatures/fields are shown individually for upward and downward temperature/field sweeps.}
     \label{fig:Phase_diagram_both_sweep_directions}
\end{figure}

The authors ascribe the irreversibility of their measured phase diagrams to an intrinsic behavior of the odd-parity state. We propose a much simpler explanation---the magnetocaloric heating and cooling of the BeCu body of the pressure cell. Decreasing the magnetic field cools the cell down via an adiabatic demagnetisation involving the copper nuclear spins, and the sample can become colder than what the thermometer reports. The effect is reversed when the magnetic field is swept up, which is why the observed $H_{\mathrm{c2}}$ values are higher for down-sweeps than for the up-sweeps [see, for example, Fig.~3(a,b) of Siddiquee et al.]. This effect becomes more pronounced at lower temperatures and higher fields, which is where the SC2 state occurs. The magnetocaloric effect of such magnitude would imply a sizable uncertainty in temperature for the critical field data in Fig.~3(d) of Ref.~\cite{siddiquee2023}, making it impossible to conclude whether the SC1-SC2 transition is observed or not. In Fig.~\ref{fig:Phase_diagram_both_sweep_directions} we show our $c$-axis critical field data at 2.50\,GPa, showing the transition fields and temperatures for sweeps in both directions. Even with the slow 0.02\,T/min sweep rate used in our measurement, a slight irreversibility between the field sweeps was unavoidable, but the mismatch is much smaller than in Ref.~\cite{siddiquee2023}, disfavoring the intrinsic origin of the phenomenon.

\begin{figure}[b]
     \includegraphics[width=\columnwidth]{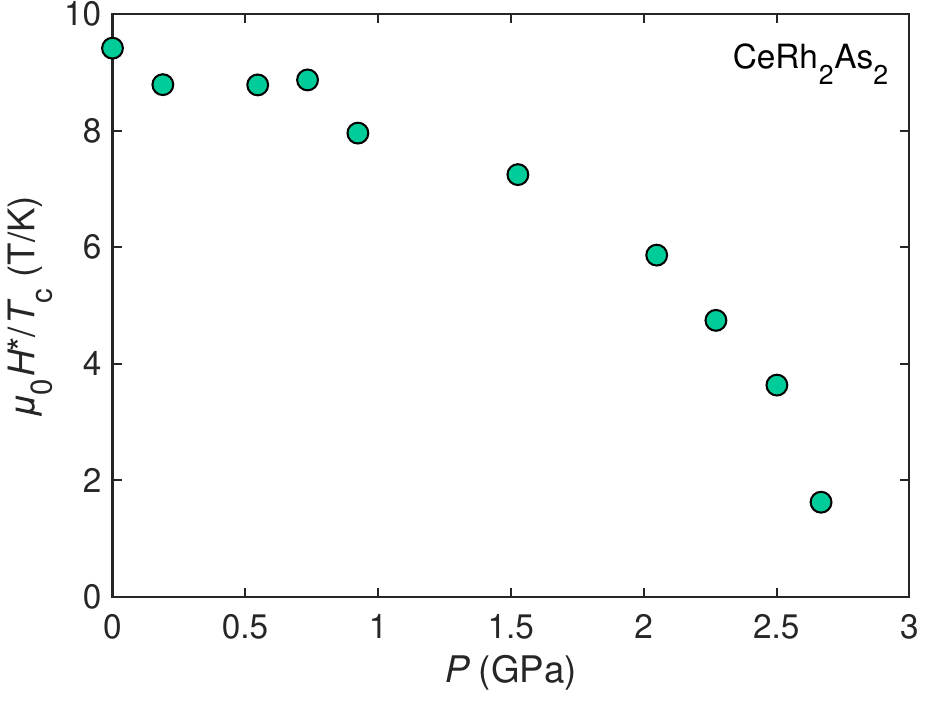}
     \caption{Pressure dependence of the ratio of the superconducting phase switching field to the superconducting critical temperature for \CRA.}
     \label{fig:Hstar_over_Tc_vs_P}
\end{figure}

3). In Fig.~2 and Fig.~3 of Ref.~\cite{siddiquee2023}, the scatter of the points along the critical field curve as well as the strong difference between upward and downward field sweeps make it impossible to determine $H^{*}$ past 2\,GPa. The attribution of the points to the even-/odd-parity phase appears arbitrary at higher pressures, as acknowledged by the authors. 
They write: "Above 2.5\,GPa the kink in $H_{\mathrm{c2}}$-$T$ is diminished, and a more detailed analysis is required to investigate whether there is still a transition from even- to odd-parity state." Furthermore, the authors state that "Above 2.5\,GPa, an interesting situation happens: while the low-field part can still fit to the expression for the even-parity state, the whole data range can also fit well to the expression for the odd-parity state. In other words, the low-field part of the data can fit to both even- and odd-parity states." This ambiguity affects the pressure dependence of $H^{*}/T_{\mathrm{c}}$ in Fig.~4(c) of Ref.~\cite{siddiquee2023}, where it is nearly constant, which drastically differs from the one found in our work, displayed in Fig.~\ref{fig:Hstar_over_Tc_vs_P} and showing a strong decrease for pressures above 2\,GPa.

4). The primary conclusion of Ref.~\cite{siddiquee2023} is that for ${P<2.5}$\,GPa "The odd-parity is suppressed faster than the even-parity state, which could be consistent with the competition between the Rashba spin-orbit coupling and the interlayer hopping in determining the parity of superconducting states of systems with local symmetry breaking". The authors propose that the local non-centrosymmetry parameter $\alpha/t_{\perp}$ is decreasing under pressure due to the increase of $t_{\perp}$, and the system approaches a state where only the SC1 phase is present. Such a deduction, which is of essential relevance for the understanding of the unconventional superconductivity in \CRA, gets very strongly questioned by the pressure dependences of $T_{\mathrm{c}}^{\mathrm{SC2}}/T_{\mathrm{c}}^{\mathrm{SC1}}$ and $H^{*}/T_{\mathrm{c}}$ found in our study. The stabilization of the SC2 state relative to SC1 is supported by the ratio $T_{\mathrm{c}}^{\mathrm{SC2}}/T_{\mathrm{c}}^{\mathrm{SC1}}$ steadily approaching unity when pressure is increased past 2\,GPa [Fig.~\ref{fig:Discussion}(d) of the main text]. A secondary indication of that is the strong reduction of $H^{*}/T_{\mathrm{c}}$ (Fig.~\ref{fig:Hstar_over_Tc_vs_P}) in the $P>2$\,GPa range, yet in Fig.~4(c) of Ref.~\cite{siddiquee2023}, the same ratio does not exhibit a pronounced pressure dependence. This discrepancy is linked to the previously mentioned ambiguity in $H^{*}$ for $P>2$\,GPa, which is the regime where we find the change to be most significant. Consequently, based on the Ref.~\cite{siddiquee2023} data alone, the authors were not able to make any firm conclusions regarding the effect of pressure on the multi-phase superconductivity in \CRA.

\bibliography{CeRh2As2_pressure_SC}
\end{document}